\documentstyle[preprint,aps]{revtex}
\begin{document}

\draft
\title{
Heat conduction in the heavy fermion superconductor UPd$_2$Al$_3$}

\author{
May Chiao, Benoit Lussier, Brett Ellman and Louis Taillefer\\
}

\address{Department of Physics,
McGill University, 3600 University Street,
Montr\'eal, Qu\'ebec, Canada H3A 2T8
}

\date{\today}

\maketitle

\begin{abstract}

We present preliminary measurements of the thermal conductivity
of the heavy
fermion superconductor UPd$_2$Al$_3$ for the
normal and superconducting
states at low temperatures.
As T$\to$0, the superconducting state is characterized by a finite
linear term, about 10\% of the normal state value, which suggests a residual
density of low energy quasiparticle states.  This agrees qualitatively with
resonant impurity scattering theories applied to exotic superconductors with
nodes in the gap structure.  Comparisons are made with theory and with other
U-based superconductors, such as URu$_2$Si$_2$ and UPt$_3$.

\end{abstract}

Heavy fermion superconductors (HFS) have drawn much attention since
their discovery in the mid-80's, yet the nature of the order parameter
remains a focus of intense research.  Discovered in 1991,
UPd$_2$Al$_3$ exhibits
co-existing antiferromagnetism and superconductivity~\cite{Geibel1993},
with the highest T$_c$ (2 K) and magnetic moment (0.85$\mu_B$/U) of all
the HFS.
NMR relaxation rates~\cite{Kyogaku}
show a power law T dependence
suggestive of a gap with a line of zeroes on the
Fermi surface.
However, the T dependence of the specific heat [1]
and the penetration depth ~\cite{Amato}
are quite close to that expected for an s-wave gap.
In this paper, we make a preliminary study
of the gap structure using thermal
conductivity.

The polycrystalline sample was grown and annealed using RF induction
heating in
ultrahigh vacuum.  We measured the thermal conductivity $\kappa$ using the
steady-state
technique, with one heater and two thermometers.
The normal state thermal conductivity $\kappa_N$ was
obtained by applying a field H$>$H$_{c2}$(0).
This sample
has a T$_c$ of 1.86 K and an RRR of 36
($\rho_0=3.8 \mu\Omega$cm).
The positive magnetoresistance is roughly 20\%
in 4.5T.

To use $\kappa$ as a probe of gap structure, it is necessary to
separate the contributions from all the heat carriers.
Let us first consider the normal state (Fig. 1).
In a field of 4.5 T, the Lorentz number
L=$\kappa\rho$/T at 100 mK is
$2.32\times 10^{-8}$ W$\Omega$K$^{-2}$, which is 0.95L$_0$
(L$_0$=$2.44\times 10^{-8}$ W$\Omega$K$^{-2}$),
in agreement
with the Wiedemann-Franz law.
In the absence of
inelastic scattering, whose effect on $\kappa$ is unlikely to
exceed 10\% below 1 K, such behaviour is expected from electrons. As a
result, we take the electronic part of $\kappa_N$/T to be
L$_0$/$\rho_0$ from 0 to 1 K.  From inelastic
neutron scattering~\cite{Petersen}, the antiferromagnetic magnons have a
short
life-time of about $3\times 10^{-13}$ s.  With the magnon
velocity,
we get a mean free path of 13 $\AA$.  Since the spin wave excitation
spectrum is gapless and linear in q, the
specific heat
c$_{mag}$ goes as T$^3$.  Then the magnon $\kappa$ can be estimated by
$\kappa_{mag}$=c$_{mag}$v$_{mag}$l$_{mag}$/3, where c$_{mag}$=25.4
Jm$^{-3}$K$^{-4}\times$T$^3$ ~\cite{Caspary},
v$_{mag}$=4.5$\times 10^3$ ms$^{-1}$ ~\cite{Petersen}
and l$_{mag}$=13 $\AA$.  At 1 K,
$\kappa_{mag}$=5.0$\times 10^{-4}$ mWcm$^{-1}$K$^{-1}$, which is
negligible compared to the data.  Furthermore,
spin-waves are not found to be affected by the
superconducting transition
\cite{Petersen}. For
these reasons, we attribute the measured $\kappa$ entirely to
phonons and electrons. Therefore, the
roughly linear increase in $\kappa_N$/T seen in Fig. 1 must be due to
phonons.
Then, using
$\kappa_{ph}=c_{ph}v_{ph}\Lambda_{ph}/3$,
where
the specific heat c$_{ph}$=8.0$\times$T$^3$ Jm$^{-3}$K$^{-4}$~\cite{Caspary}
and the average sound velocity
$v_{ph}$=5.73$\times 10^3$ ms$^{-1}$ (from
elastic constants~\cite{Modler}),
we estimate the phonon mean free path
$\Lambda_{ph}$ to be 50 $\mu$m
at 0.6K.
By 175 mK it has increased to 160 $\mu$m,
roughly 1/4 the sample size.
These are surprisingly long phonon mean free paths for a metal.  Assuming
the dominant scattering of phonons to come from
grain boundaries and electrons,
we can estimate the boundary scattering rate
B and the e-ph coupling strength E from:

\begin{equation}
\kappa_{ph} = \frac{k_B^4 T^3}{2 \pi^2 \hbar^3 v_{ph}} \int _0 ^{\infty}
 {dx \frac {x^4 e^x (e^x-1)^{-2}}{B+ExT}}
\end{equation}

\noindent
A fit to $\kappa_N(T)$-$\kappa_N(0)$ using Eq.(1) yields
B=1.5$\times 10^6$ s$^{-1}$
and E=$5.9\times 10^6$ K$^{-1}$s$^{-1}$, which means
that electron scattering dominates at all but the lowest temperatures.
This value of E is somewhat less than
in URu$_2$Si$_2$ ($1.5\times 10^7$)~\cite{Kamran}, and much less
than in Nb ($2\times 10^9$) \cite{Kes} and V($3\times 10^9$).
This anomalously weak e-ph coupling
is the most unusual feature of the normal state $\kappa$ in UPd$_2$Al$_3$.

Let us now turn to the superconducting state data, shown as the lower curve
in Fig. 1. The two main observations are: 1)
the existence of a substantial intercept in $\kappa/T$, 2)
the similarity in the slopes of $\kappa/T$ and $\kappa_N/T$ at low T.
A smooth extrapolation of the data to T=0 gives a limiting
$\kappa$/T=0.6-0.8 mWK$^{-2}$cm$^{-1}$, about 10\% of
the normal state value L$_0$/$\rho_0$, likely due to
residual quasiparticle excitations.
Now, as seen in BCS superconductors such as Nb \cite{Kes},
the large
reduction in the number of quasiparticles available for scattering
phonons at low T should cause a major increase in $\Lambda_{ph}$.
If the rise in $\kappa_N/T$ is due to phonons scattering
off electrons, one would expect the slope of $\kappa$/T
in the superconducting state to
increase dramatically.  It does not.  In fact, it
seems that the non-electronic contribution to $\kappa_N/T$ is
unaffected by superconductivity.
Whether other scattering mechanisms play a
role requires further study. At this stage, we adopt the following
simple procedure:
the electronic thermal conductivity $\kappa_e/T$ in the superconducting state
is obtained by subtracting from the H=0 data in Fig. 1 the slope
of $\kappa_N/T$. The result is shown in Fig. 2,
normalized by L$_0$/$\rho_0$.

A finite limiting value for $\kappa_e/T$
is one of the major predictions of current theories of transport in
unconventional superconducting states (see [9,10], and references
therein).
A missing U-atom acts
as a Kondo impurity in a
compensated lattice, causing multiple scattering and large
phase shifts
$\delta_0$=$\pi/2$.  Within such a resonant impurity
scattering model, a
line of nodes in the gap can
give a finite intercept in $\kappa/T$
vs T~\cite{Norman 96}.  Such a linear term arises from node smearing due to
impurities, which leaves part of the Fermi surface gapless, with a
residual density of low energy quasiparticle states.  The simplest candidate
gap structures are based on a spherical Fermi surface [9,11].  A polar gap
has a line of nodes along the
equator, an axial gap point nodes at the poles
and the hybrid gaps both line and point
nodes with the gap
approaching zero at the poles either with a linear (type I)
or quadratic (type II) k-dependence.  Calculations such as those performed
for UPt$_3$ [9] give the curves shown
in Fig. 2~\cite{Norman}. The
unitary limit is assumed, inelastic scattering neglected and the
impurity
scattering rate taken to be
$\Gamma_0 \equiv \frac{1}{2 \tau}$=0.3T$_c$~\cite{Norman},
in agreement with the value determined from
de Haas-van Alphen measurements~\cite{Inada}.
The comparison in Fig. 2 shows that the theory predicts the right
magnitude for the residual linear term, which is independent of
the way we treat the phonons. This is not true for the T dependence, thus the
discrepancy seen in Fig. 2 must be taken with caution.

It is interesting to compare with other HFS. In UPt$_3$ single
crystals, there is no evidence for a finite intercept
{}~\cite{Lussier}. This may be due to a lower impurity scattering rate, in
accordance with the lower
residual resistivities (0.23 and 0.61 $\mu\Omega$cm).
However, since T$_c$ is 0.5 K, estimates yield
$\Gamma_0$=0.1-0.2 T$_c$ ~\cite{Lussier},
while a fit to UPt$_3$ data would have
$\Gamma_0$=0.05T$_c$ or less [9].  As for URu$_2$Si$_2$, a large residual
$\kappa/T$ -- approximately 30\% of the normal state value --
was observed for a crystal with $\rho_0$=9.5 $\mu\Omega$cm [7].
Roughly speaking,
this makes sense within the theory since
$\rho_0$ is 3 times greater than in our UPd$_2$Al$_3$ sample.
Finally, we point out that in URu$_2$Si$_2$,
the phonon contribution is more easily explained.  The slope in
$\kappa/T$ in the superconducting state at low T
is roughly 4 times steeper
than in the normal state, consistent with the idea that phonons conduct
better due to the loss of electrons, their main
scatterers.  Again, it is
puzzling that in UPd$_2$Al$_3$, with a comparable e-ph coupling,
the
loss of electrons appears to have no affect on the phonons.

In conclusion, although the identification of the order parameter remains an
issue of debate in HF superconductors, the possibilities are being
constrained.  In this work on UPd$_2$Al$_3$,
we see a clear gapless behaviour
which suggests the presence of a line of nodes, consistent with NMR
results and calculations based on resonant impurity scattering.

We are grateful to M.R. Norman for his calculations.
This work was funded by NSERC of Canada and FCAR of Qu\'{e}bec.
We acknowledge the support of the Alexander McFee Foundation (M.C.),
the Canadian Institute for Advanced Research (L.T.) and the A.P. Sloan
Foundation (L.T.).

\begin{figure}
\caption{
Thermal conductivity of UPd$_2$Al$_3$, divided by temperature,
for H=0 (solid circles) and H=4.5T$>$H$_{c2}$(0) (open circles).
}
\label{fig1}
\end{figure}

\begin{figure}
\caption{
Normalized electronic thermal conductivity
(see text).  The data (points) are compared with resonant
impurity scattering calculations ([9])
using spherical
harmonics for 3 gaps with a line node:
polar (solid line), hybrid I (dashed line) and hybrid II (dotted line) [11].
}
\label{fig2}
\end{figure}

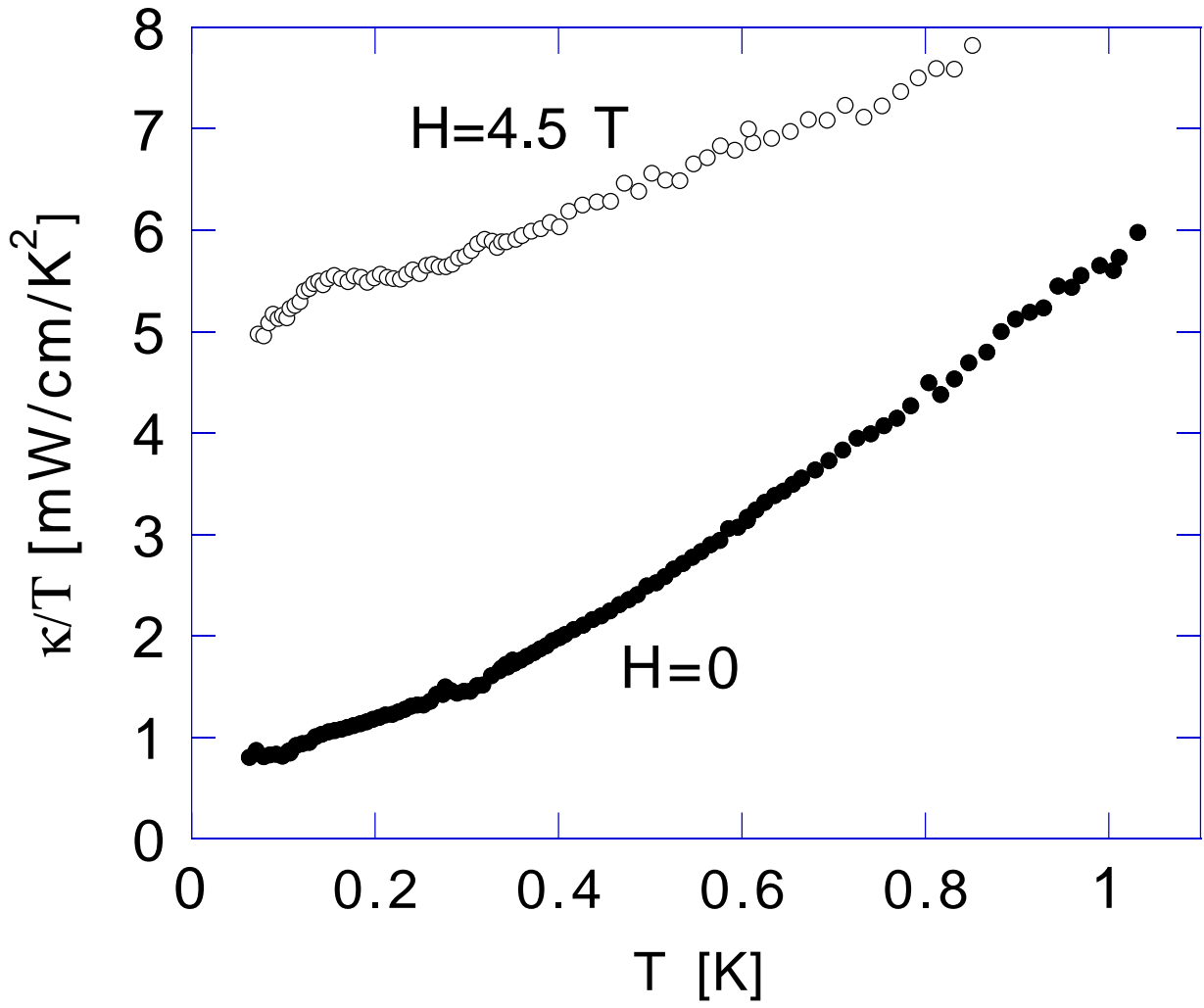
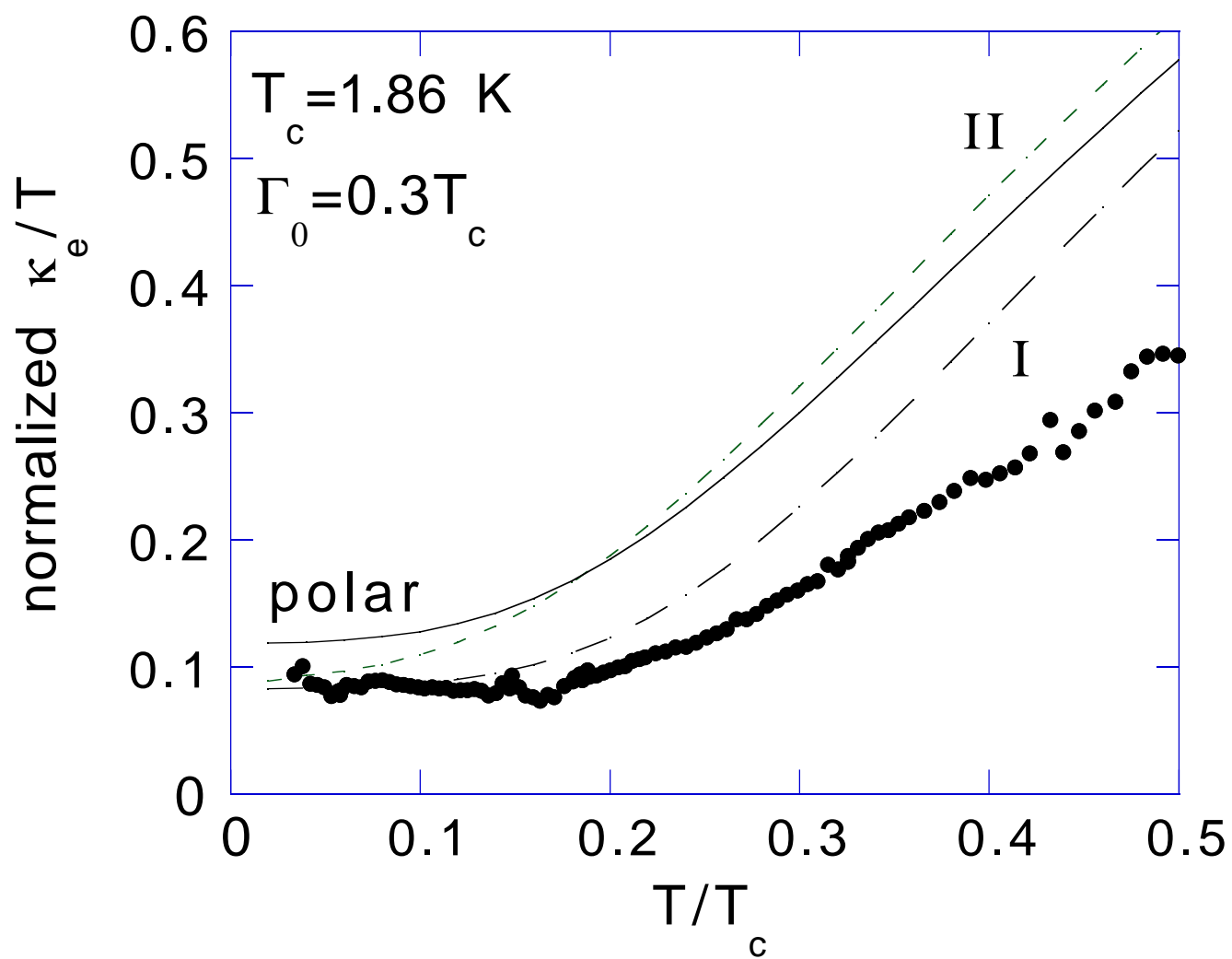
\end{document}